
\magnification 1200
\baselineskip 20pt
\vskip 1cm

\centerline{\bf Resolution of a ``Retrodiction Paradox''}
\vskip 2cm

\centerline{Yakir Aharonov$^{a,b}$ and Lev Vaidman$^a$}

\vskip 1cm

\centerline{\it $^a$  School of Physics and Astronomy}
\centerline{\it Raymond and Beverly Sackler Faculty of Exact Sciences}
\centerline{\it Tel-Aviv University, Tel-Aviv, 69978 ISRAEL}

\vskip .5cm

\centerline{\it $^b$ Physics Department, University of South Carolina}
\centerline{\it Columbia, South Carolina 29208, U.S.A.}

\vskip 2.5cm
\centerline{ Abstract}
\vskip .5cm

It is shown that the ``retrodiction paradox'' recently introduced by Peres
arises not because of the fallacy of the time-symmetric approach as he
claimed, but due to an inappropriate usage of retrodiction.

\vfill
\break

In a recent Letter$^1$ Peres claimed that our time-symmetric approach to
quantum theory,$^{2,3}$ in which retrodictions and predictions are on equal
footing, leads to a paradox. We shall show that the paradox arises due to
a particular  usage of retrodiction by Peres, and no such paradox arises in our
approach.

A trivial time asymmetry of a quantum measurements is illustrated by the
following example. Assume that the $ x$ component of the spin of a spin-1/2
particle was measured at time $t$, and was found to be $\sigma_x =1$. While
there is a symmetry regarding prediction and retrodiction for the result of
measuring $\sigma_x$ after or before the time $t$ (in both cases we are
certain that $\sigma_x =1$), there is an asymmetry regarding the results of
measuring $\sigma_y$. We can predict equal probabilities for each outcome,
$\sigma_y = \pm 1$, of a measurement performed after the time $t$, but we
cannot claim the same for the result of a measurement of $\sigma_y$
performed before the time $t$. The difference arises from  time
asymmetric basic preconception, namely the usual assumption that there is
no ``boundary condition'' in the future, but there is a boundary condition
in the past: the state in which the particle was prepared before the time
$t$. For more details see Sec. II of Ref. 2.

The time symmetric approach is applicable in a situation when we consider
a quantum system at a time between two complete measurements which yield
{\it two} boundary conditions. Then, the time symmetry of the formalism of
quantum theory$^4$ together with the symmetry of having  boundary
conditions {\it both} in the past and in the future  allow
us to apply our time symmetric approach. We describe the system by
a two-state vector consisting of the state evolving from the measurement in
the   past and the backward evolving state evolving from the complete
measurement in the future. We  are making inferences about  the results of
measurements (if performed) at the time $t$ using  both prediction and
retrodiction. The situation considered by Peres  essentially falls into
this latter category,
and therefore we can apply our approach for its analysis.

Peres considers an ensemble of pairs of separated spin-1/2 particles,
initially (before time $t$) prepared in a singlet spin state. This
pre-selected ensemble is divided into four sub-ensembles according to
the randomly chosen measurement performed after the time $t$. Peres
focuses on the sub-ensemble in which $\sigma_{Ax}$ was
measured. Although in this case no complete measurement has been
performed on the pair of particles, we are in the position to infer,
from the measurement before the time $t$ (the singlet state) and the
measurement after the time $t$ (known $x$ component of the spin of
particle $A$), the result of a measurement of $\sigma_x$ of particle
$B$. From the correct inference that the outcome of the measurement of
$\sigma_{Bx}$ at time $t$ is known with certainty Peres concludes that
particle $B$ was, from the beginning, in one of the eigenstates:
either $|{\uparrow_x}\rangle$ or $|{\downarrow_x}\rangle$.  In our
approach, this conclusion is incorrect. Initially and at time $t$, the
pair of particles $A$ and $B$ is in a pure forward-evolving singlet
state and no measurement at a time later than $t$ can change
this. Therefore, particle $B$ can not be described by a pure
forward-evolving quantum state.

  Peres encounters a retrodiction paradox because he uses the standard
approach of a single forward-evolving state. In this approach the only
possibility in which the result of a measurement can be known with
certainty is when the system is in an eigenstate of the measured
variable. In the two-state vector approach, when we use both
prediction and retrodiction, there are many situations in which
neither the forward evolving state nor the backward evolving state are
eigenstates of the measured variable but, nevertheless, the result of
the measurement at time $t$ (if performed) is known with
certainty. See Refs. 5,6 for relevant the examples.

 Retrodiction is not ``rewriting history'' as Peres claims. It comes
not instead of but in addition to prediction. The combination of the
two allows Lorentz-invariant time symmetric quantum description which
is something more than just a mathematical tool for calculating
probabilities.

\vskip 1.5 true cm
\noindent
 REFERENCES
\vskip .5 true cm

\noindent
1. A. Peres, {\it Phys.  Lett.} {\bf A 194}, 21 (1994).\hfil
\break
2.  Y. Aharonov and L. Vaidman, {\it Phys.  Rev.}   {\bf A 41}, 11 (1990).\hfil
\break
3.  Y. Aharonov and L. Vaidman,  {\it J. Phys.} {\bf A 24}, 2315 (1991).\hfil
\break
4. Y. Aharonov, Bergmann,
and J.L.  Lebowitz, {\it Phys.  Rev. } {\bf B 134}, 1410 (1964).\hfil
\break
5. L. Vaidman, Y. Aharonov, and Albert, {\it Phys.  Rev.  Lett.} {\bf 58}, 1385
 (1987).
\hfil
\break
6. L. Vaidman, {\it Phys.  Rev.  Lett.} {\bf 70}, 3369 (1993).

\end